\newcommand{\ep}{\epsilon}
\newcommand{\eps}{\varepsilon}  
\newcommand{\de}{\delta} 
\newcommand{\vphi}{\varphi} 
\newcommand{\cl}{\mbox{\rm cl}}

\newcommand{\C}{{\Bbb C}}\newcommand{\R}{{\Bbb R}}\newcommand{\N}{{\Bbb N}}  
\newcommand{\pa}{\partial} 
 \newcommand{\G}{{\cal G}} 
\newcommand{\DD}{{\cal D}}\newcommand{\E}{{\cal E}}

\newcommand{\fr}{\frac{1}} 
\newcommand{\nn}{\nonumber} 
\newcommand{\ba}{\begin{array}}\newcommand{\ea}{\end{array}} 
\newcommand{\beq}{ \begin{equation} }\newcommand{\eeq}{ \end{equation} } 
\newcommand{\bea}{\begin{eqnarray}}\newcommand{\eea}{\end{eqnarray}} 
\newcommand{\beas}{\begin{eqnarray*}}\newcommand{\eeas}{\end{eqnarray*}} 
\newcommand{\beqn}{ \begin{equation*} } 
\newcommand{\lb}{\label} 
 
\newtheorem{definition}{Definition} 
\newtheorem{lemma}{Lemma} 
\newtheorem{theorem}{Theorem} 
 
\newtheorem{prf}{{\it Proof}} 
\newcommand{\pr}{{\bf Proof. }} 
\newcommand{\epr}{\hspace*{\fill}$\Box$} 
 
\documentstyle[aps,amsfonts]{revtex} 
\title{A rigorous solution concept for geodesic and geodesic deviation 
equations in impulsive gravitational waves} 
\author{M. Kunzinger 
	\footnote{Electronic mail: Michael.Kunzinger@univie.ac.at}, 
	 R. Steinbauer 
	 \footnote{Electronic mail: stein@doppler.thp.univie.ac.at}} 
\address{Department of Mathematics, University of Vienna,  
	Strudlhofg.~4\\ 
	Institute for Theoretical Physics, University of Vienna, 
        Boltzmanng.~5 
	\\ A-1090 Wien, Austria} 
 
\begin{document} 
\maketitle 
\thispagestyle{empty}            
 
\begin{abstract} The geodesic as well as the geodesic deviation equation for 
impulsive gravitational waves involve highly singular products of distributions 
$(\theta\de$, $\theta^2\de$, $\de^2$). A solution concept for these equations 
based on embedding the distributional metric into the Colombeau algebra of 
generalized functions is presented. Using a universal regularization procedure 
we prove existence and uniqueness results and calculate the distributional 
limits of these solutions explicitly. The obtained limits are regularization 
independent and display the physically expected behavior. 

{\em PACS-numbers: }04.20.Cv, 04.20.-q, 02.20.Hq, 04.30.-w  

\end{abstract} 
 
\rightline{UWThPh -- 1998 -- 25} 
\section{Introduction} 
 
 
Impulsive pp-waves (plane fronted gravitational waves with parallel rays) can 
be described by a metric of the form~\cite{penrose}
\beq\lb{me}ds^2\,=\,\de(u)\,f(x,y)du^2-du\,dv+dx^2+dy^2,\eeq
where $(u,v)$ and $(x,y)$ is a pair of null and (transverse) Cartesian
coordinates respectively, and $f$ denotes the profile function subject to the
field equations. Hence the spacetime is flat everywhere except for 
the null hypersurface $u=0$, where it has a $\de$-like pulse modelling 
a gravitational shock wave. Such geometries arise most naturally as 
ultrarelativistic limits of boosted black hole spacetimes of the Kerr-Newman 
family (as shown by various authors~\cite{as,bn,ls}) and multipole solutions of
the Weyl family~\cite{pg}. Also they play an important role  
in particle scattering at the Planck scale (see~\cite{vv} and references  
therein). 
 
There have been intrinsic descriptions of impulsive pp-waves, viz. by
Penrose~\cite{penrose} and by Dray and t'Hooft~\cite{dt}, which essentially 
consist in glueing together two copies of Minkowski spacetime with a warp  
across the null hypersurface $u=0$. 
Penrose also introduced a different coordinate system in which the components 
of the metric tensor are actually continuous. 
However, the transformation relating the coordinates used in (\ref{me}) to
these new ones is discontinuous (for the general form of the transformation
see~\cite{aib}) and therefore --strictly speaking-- the differential structure
of  the manifold is changed.  In this paper we stick to the original
distributional form of the metric,   motivated by the fact that physically,
i.e. in the ultrarelativistic limit,   the spacetime arises that way (cf. the
approaches of~\cite{ba,fvp}). For recent work on pp-waves using the continuous
form of the metric see~\cite{vp}.
 
We describe the geometry of impulsive pp-waves entirely in the 
distributional picture using the framework of Colombeau's generalized  
functions, thereby generalizing previous work~\cite{geo}.  
As discussed  there in detail, the geodesic as 
well as the geodesic deviation equation for impulsive pp-waves  
involve formally ill-defined products of 
distributions, due to the nonlinearity of the equations and the presence of the 
Dirac $\de$-function in the space time metric. However, as was also shown 
in~\cite{geo}, one can overcome these difficulties using a careful 
regularization procedure which, while mathematically sound, corresponds  
to the physical idea of viewing the impulsive wave as the limiting case of a  
sandwich wave of ever decreasing support but constant (integrated) strength.  
More precisely, regularizing the $\de$-distribution by a ``model $\de$-net''  
(i.e., a net $\rho_\ep(x):=\ep^{-1}\rho(x\ep^{-1})$, where $\rho$ is a smooth  
function with support contained in the interval $[-1,1]$ satisfying  
$\int\rho=1$), it was shown that the solutions to the smoothened equations  
possess regularization independent weak limits.  
These distributional ``solutions'' fit perfectly into the physically expected  
picture showing that the geometry of impulsive pp-waves can be described  
consistently using the distributional form of the metric. 
The reliability of the results is guaranteed by making use of regularization 
techniques instead of introducing ``multiplication rules'' into Schwartz linear 
disribution theory (cf. the discussion at the end of Sec.\,2 in~\cite{geo}
or~\cite{hajek} for general remarks).

However, the ``solutions'' obtained by this naive regularization procedure  
exhibit a mathematically highly unsatisfactory feature. {\em They 
do not obey the original distributional equations} (unless, again, one is 
willing to impose certain ``multiplication rules''), as is common  
to such situations.  
Hence --strictly speaking-- this approach does not provide a reasonable
solution concept for the equations under consideration. 
Such a notion {\em is} available in the nonlinear theory of  
generalized functions~\cite{co1,co2,mo} due to J.\,F. Colombeau, where one has 
--loosely speaking-- a rigorous system of bookkeeping on the regularizing 
sequences.  
Recently Hermann and Oberguggenberger~\cite{moODE} (see also~\cite{kunzDISS}) 
studied systems of singular, nonlinear ODEs in the Colombeau algebra.  
In this work we are going to use
similar techniques to treat the geodesic and geodesic deviation equation for 
impulsive pp-waves in the Colombeau algebra. Despite the nonlinearities
involved in these equations (which in principle could lead to trapping,
blow--up or reflection of solutions at the shock, cf.~\cite{moODE}) we are
able to prove existence and uniqueness of geodesics crossing the shock
hypersurface.
We derive the (regularization independent) distributional limits of these
solutions, making use of the  notion of association (see Sec.~\ref{ma} below)
in the algebra,   thereby significantly generalizing the results
of~\cite{geo}. In  particular, the regularization of the $\de$-like wave
profile will no longer be restricted to a ``model $\de$-net'' but belong to
the largest ``reasonable'' class  (cf.  Definition~\ref{gendelta}  below).
Moreover, note that the regularization idependence of the results has the
following important physical consequence: in the impulsive limit the geodesics
are totally independent of the particular shape of the sandwich wave. Hence
the impulsive wave ``totally forgets its seed'' (cf. also the results
in~\cite{pv2}).  

Finally, we discuss the 
case of a nonsmooth wave profile $f$ and give an outlook to current research
which allows to fit our  previous calculations into a manifestly covariant
concept of Colomebeau  algebras on manifolds.

\section{Mathematical Framework}\label{ma} 
A  framework that  allows  consistent  treatment  of  nonlinear 
operations  with  distributions and at the same time offers a 
well-developed   theory   of  (linear  and  nonlinear)  partial 
differential  equations  is  provided  by Colombeau's theory of 
algebras    of    generalized    functions (cf. e.g.  \cite{co1,co2,mo,AB}).
To  begin  with,  we give a short description of the algebra we 
are going to use in the sequel. Let 
\beas 
 {\cal  A}_0(\R^n)  &=&  \{\vphi  \in  {\cal D}(\R^n) \, : \, \int 
\vphi(x)\,dx = 1\}  \\ 
 {\cal  A}_q(\R^n)  &=&  \{\vphi  \in  {\cal  A}_0(\R^n) \, : \, \int 
\vphi(x)  x^\alpha  \,dx  =  0\,  ,  \,  1\le  |\alpha| \le q\} 
\,\,\,(q \in \N) 
\eeas 
and set (for any $\Omega\subseteq \R^n$ open)  
\beas 
{\cal E}(\Omega) &=& \{R: {\cal A}_0(\R^n) \times \Omega \to \C \, :\, 
  x\to R(\vphi,x) \in {\cal C}^\infty(\Omega)\, \forall \vphi \in 
  {\cal A}_0(\R^n) \} \\ 
{\cal E}_M(\Omega) &=& \{u\in {\cal E}(\Omega) :  
  \forall K\subset\subset \Omega \ \forall \alpha\in \N_0^n \ \exists p\in 
  \N_0 \ \forall \varphi\in {\cal A}_p(\R^n)  \\ 
&&\quad\exists  c>0 \ \exists \eta>0 \ \sup\limits_{x\in K} | \partial^\alpha 
  u(\varphi_\eps,x) | \le c \eps^{-p} \ (0<\eps<\eta)\}  \\ 
{\cal N}(\Omega) &=& \{u\in {\cal E}(\Omega) :  
  \forall K\subset\subset \Omega \ \forall \alpha\in \N_0^n \ \exists p\in 
  \N_0 \ \exists \gamma\in \Gamma \ \forall q\ge p  \\ 
&&\quad\forall \varphi\in {\cal A}_q(\R^n)\   
  \exists  c>0 \ \exists \eta>0 \ \sup\limits_{x\in K} | \partial^\alpha 
  u(\varphi_\eps,x) | \le c \eps^{\gamma(q)-p} \ (0<\eps<\eta)\}\,\,,  
\eeas 
where $\Gamma = \{ \gamma:\N_0\to \R_+ : \gamma \mbox{ strictly increasing, } 
\lim\limits_{n\to \infty} \gamma(n) = \infty \}$. 
Derivation $\partial^\alpha$ is carried out with respect to $x$, 
while the  $\vphi$  are treated as parameters. 
Also,    for    $\vphi\in    \DD(\R^n)$, 
$\vphi_\eps(x) = \eps^{-n} \vphi(\frac{x}{\eps})$. Note that  
$\vphi_\eps \to \delta$ in $\DD'(\R^n)$.  
 
Elements of ${\cal E}_M(\Omega)$ are called of {\em moderate growth}.  
With pointwise operations ${\cal E}_M(\Omega)$ 
is a differential algebra and ${\cal N}(\Omega)$ is an ideal in 
${\cal E}_M(\Omega)$. The quotient algebra 
\[ 
{\cal G}(\Omega) = {\cal E}_M(\Omega)/{\cal N}(\Omega) 
\] 
is  called  the  {\em Colombeau  algebra} over $\Omega\subseteq\R^n$. 
Elements   of   ${\cal   G}(\Omega)$   will   be   denoted   by 
$R=\cl[(R(\vphi,\,.\,))_{\vphi\in {\cal A}_0}]$ where 
$(R(\vphi,\,.\,))_{\vphi\in   {\cal   A}_0}$   is   an  arbitrary 
representative  of $R$ (again emphasizing the fact that the $\vphi$'s 
are viewed as parameters).  
 
For $\Omega=\R^n$ the map  
\begin{eqnarray*} 
& \iota:{\cal E}'(\Omega) \to {\cal G}(\Omega)    & \\ 
& w \to \cl [(w\ast \varphi)_{\varphi\in {\cal A}_0}] & 
\end{eqnarray*} 
(where  $\ast$  denotes  convolution)  is  a  linear  embedding 
commuting   with   partial  derivatives and coinciding with the 
identical embedding $f\to \cl[(f)_{\vphi\in {\cal A}_0}]$ on  
${\cal D}(\R^n)$.  
 
${\cal G}$ is a 
fine  sheaf  of  differential algebras on $\R^n$ and there is a 
unique sheaf morphism $\widehat{\iota}: {\cal D}' \to {\cal G}$  
coinciding  with $\iota$ on every ${\cal E}'(\Omega)$ and rendering 
${\cal C}^\infty(\Omega)$ a  faithful  subalgebra  of  
${\cal G}(\Omega)$. From the 
definitions  it  is  clear  that any element of $\G(\Omega)$ is 
uniquely  determined  by  the  values  of any representative on 
$\vphi_\eps$  for  $\vphi\in  {\cal  A}_p$ with $p$ arbitrarily 
large  and  $\eps$ arbitrarily small (i.e. by its  `germ'), a fact 
that  turns out to be very helpful e.g. in constructing solutions  
to differential equations in $\G$. 
 
Inserting  points into elements of $\G(\R^n)$ gives elements of 
the  ring of generalized numbers $\overline{\C}(n)$, defined as 
$\overline{\C}(n)={\cal E}(n)/{\cal N}(n)$, where 
\beas 
{\cal E}(n) &=& \{u: {\cal A}_0(\R^n)\to \C : \exists p\in \N_0 
\ \forall \varphi\in {\cal A}_p(\R^n) \ \exists c>0 \ \exists \eta >0 \\ 
 &&\quad |u(\varphi_\eps)|\le c \eps^{-p} \ (0<\eps<\eta) \}\\ 
{\cal N}(n) &=& \{u: {\cal A}_0(\R^n)\to \C : \exists p\in \N_0 
  \ \exists \gamma\in\Gamma \ \forall q\ge p 
  \ \forall \varphi\in {\cal A}_q(\R^n)  \\ 
&&\quad \exists c>0 \ \exists \eta >0  
  \ |u(\varphi_\eps)|\le c \eps^{\gamma(q)-p} \ (0<\eps<\eta)\}  
\eeas 
Thus  elements  of $\G(\R^n)$ take values in $\overline{\C}(n)$. 
Explicit  dependence  of  the  ring  of constants on $n$ can be 
avoided by a more refined construction of the sets ${\cal A}_q$ 
in   the   definition   of   $\G$   (see  \cite{AB}).  Clearly, 
$\C\hookrightarrow \overline{\C}$ via the canonical embedding 
$c \to \cl[(c)_{\vphi\in {\cal A}_0}]$.  
 
Componentwise insertion  
of $R\in \G$  into  a smooth function $f$ yields a well defined  
element  $f(R)$ of $\G$ if $f$ is {\em slowly increasing}, i.e. 
if  all  derivatives of $f$ are polynomially bounded. Moreover, 
if  $R$  is  {\em  locally  bounded,}  i.e.  if  it  possesses a 
representative   such  that  $R(\vphi_\eps,\,.\,)$  is  bounded 
uniformly  in $\eps$ on compact sets (for $\vphi\in {\cal A}_p( 
\R^n)$, $p$ large) then $f\circ R$ exists for any smooth $f$. 
 
Finally, 
we mention the notion of {\em association} in ${\cal G}(\Omega)$: 
$R_1$,  $R_2$  $\in  {\cal G}(\Omega)$ are called associated to 
each other ($R_1\approx R_2$) 
if there exists some $p\in \N$ such that  
$R_1(\vphi_\eps, \,.\,) - R_2(\vphi_\eps, \,.\,)\to 0$ in  
${\cal    D}'(\Omega)$  as $\eps\to 0$ for all $\vphi \in {\cal 
A}_p(\R^n)$. In particular, if $R_2\in 
{\cal  D}'(\Omega)$ then $R_2$ is called the macroscopic aspect 
(or  {\em distributional  shadow})  of  $R_1$.  Equality  in  
$\DD'$ is 
reflected  as  equality  in  the sense of association in ${\cal 
G}$, while equality in ${\cal G}$ is a stricter concept (for 
example, all  powers  of the Heaviside function are distinct in 
the  Colombeau  algebra  although they are associated with each 
other). 
\section{Exact  Solutions  of  Geodesic and Geodesic Deviation Equations  
}\label{es} 
As in \cite{geo} we consider the impulsive pp-wave metric 
\beq \label{metric} 
ds^2\,=\,f(x^i)\,\de(u)\,du^2-du\,dv+(dx^i)^2\,\,,  
\eeq 
where $f$ is a smooth function of the transverse coordinates 
$x^i$ ($i=1,2$). Our aim is to 
derive  solutions  to  the  corresponding geodesic and geodesic 
deviation  equations  in  the  Colombeau  algebra.   
 
The general 
strategy for solving differential equations in $\G$ is to embed  
singularities (in our case: $\de$) into $\G$ which amounts to a 
regularization  and  then  solve  the corresponding regularized 
equations. In order to obtain general results we are therefore  
interested    in  imposing  as  few restrictions as possible on 
the regularization of $\de$. The largest ``reasonable'' class of  
smooth\footnote{Note that since ${\cal D}$ is dense in $L^1$ 
practically even discontinuous regularizations (eg. boxes) are included.}
regularizations  of $\de$ is given by nets 
$(\rho_\eps)_{\eps\in (0,1)}$ of smooth functions $\rho_\eps$ 
satisfying:
\bea 
\mbox{(a)} 
&\quad&   \mbox{\rm   supp}(\rho_\eps)  \to  \{0\}  \quad (\eps\to  0)\,\,, 
\nn\\ 
\mbox{(b)} 
&\quad& \int \rho_\eps (x) \,dx \to 1 \quad (\eps\to 0)\,\,\mbox{ and} \nn\\ 
\mbox{(c)} 
&\quad& \exists \eta>0\,\,\exists C\ge 0:  \int |\rho_\eps (x)| \,dx  
\le C \,\, \forall  \eps\in (0,\eta) \nn 
\eea 
(cf. the definition of {\em strict delta  nets}  in \cite{mo},   
ch.  2.7). Obviously  any  such net converges to $\delta$ in  
distributions as $\eps\to 0$. 
To simplify notations it is often convenient to replace (a) 
by 
$$ \mbox{(a')}\quad 
\mbox{\rm  supp}(\rho_\eps)  \subseteq  [-\eps,\eps] \;\;\;\; 
\forall \eps\in (0,1). 
$$ 
This motivates the following (cf. \cite{moODE}) 
 
\begin{definition} \label{gendelta} 
A  {\em generalized  delta  function} is an element $D$ of $\G(\R^n)$ 
possessing a representative $(D(\vphi,\,.\,))_{\vphi\in {\cal A}_0}$ 
such that $\exists p\in \N_0 \,\, \forall \vphi\in {\cal A}_p(\R^n)\, 
\exists \eta=\eta(\vphi)>0:$ 
\bea 
\mbox{\rm (i)} 
&\quad&  
 \mbox{\rm supp}(D(\vphi_\eps,\,.\,)) \subseteq [-\eps,\eps] 
\,\,\,\forall \eps\in (0,\eta)\nn\\  
\mbox{\rm (ii)} 
&\quad&    \int   D(\vphi_\eps,x)   \,dx   \to   1   \,\,   (\eps\to 
0)\nn\\ 
\mbox{\rm (iii)} 
&\quad&\exists C=C(\vphi)>0  
\mbox{ such that } \int |D(\vphi_\eps,x)| \,dx \le C \;\;  
\forall \eps\in (0,\eta)\nn 
\eea 
\end{definition} 
The  canonical  embedding  $R=\iota(\de)$  of course falls  into 
this    class    but   clearly there are many generalized delta 
functions  that  do  not  correspond  to  any  distribution via 
$\iota$.   Moreover,   every   generalized  delta  function  is 
associated  to  $\delta$,  i.e. all generalized delta functions 
equal  $\delta$  on  the distributional level. In a sense, they 
may  be  viewed  as    `delta distributions with a more refined 
microstructure'  (fixing the additional nonlinear properties of 
the singularity). 

Again, condition  (i)  in definition \ref{gendelta} has  
been chosen  in  order  to avoid technicalities in the proofs of  
the following  results,  which,  however,  remain  true  if  (i)  
is replaced by  
$$  \mbox{(i')}\quad 
\mbox{\rm supp}(R(\vphi_\eps,\,.\,)) \to \{0\} \;\;\;\; (\eps\to 0) 
$$ 
Finally, we need the  following  technical  preparation  (which   
is  actually  a generalization of appendix A of \cite{geo}). 
 
\begin{lemma} \label{ro} 
Let  $g:\,\R^n\to\R^n$, $h:\,\R\to\R^n$ smooth and let  
$(\rho_\eps)_{\eps\in (0,1)}$ be a net of smooth func\-tions   
satisfying       (a') and   (c).   For   any 
$x_0,\dot{x_0}\in \R^n$  and any $\eps\in (0,1)$ consider the system 
\bea 
 \ddot{x}_\eps(t) &=& g(x_\eps(t))\rho_\eps(t) + h(t) \nonumber\\ 
 x_\eps(-1) &=& x_0  \label{ode}\\ 
 \dot{x}_\eps(-1) &=& \dot{x}_0 \nonumber 
\eea 
Let $b>0$, $M=\int_{-1}^1\,\int_{-1}^s |h(r)|\,drds$, 
$I=\{x\in \R^n:\, |x-x_0|\le b + |\dot{x}_0|+M\}$ and 
$\alpha=\min\left\{ \frac{b}{C\|g\|_{_{L^\infty(I)}}+|\dot{x}_0|}, 
\frac{1}{2LC},1\right\}$  with  $L$ a Lipschitz constant for $g$ 
on   $I$.   Then   (\ref{ode})   has   a   unique  solution  on 
$J_\eps=[-1,\alpha-\eps]$. Consequently, for $\eps$ sufficiently 
small  $x_\eps$  is  globally  defined  and  both  $x_\eps$ and 
$\dot{x}_\eps$  are  bounded,  uniformly  in $\eps$, on compact 
sets. 
\end{lemma} 
 
\pr The operator $f\to Af$, 
\[ 
Af(t)=x_0 + \dot{x}_0 (t+1) + \int\limits_{-1}^t\int\limits_{-1}^s 
g(f(r))\rho_\eps(r) \,drds + \int\limits_{-1}^t\int\limits_{-1}^s h(r)\,drds 
\] 
is a contraction on the complete metric space 
\[ 
\{f\in C(J_\eps,\R^n):\, |f(t)-x_0|\le b+M+|\dot{x}_0|\} 
\] 
\epr \medskip\\ 
Let  us  now  turn  to the geodesic equation for the 
pp-wave metric (\ref{metric}). Using $u$ as an affine parameter (which excludes
trivial geodesics parallel to the shock)
we obtain (cf. \cite{geo}):\\ 
\parbox{12cm}{ 
\beas 
     \!\ddot v(u)&=&f(x^i(u))\,\dot\de(u) 
              \,+\,2\,\pa_i\,f(x^i(u))\,\dot x^i\,(u)\,\de(u)\\ 
     \ddot x^i(u)&=&\frac{1}{2}\,\pa_i\,f(x^i(u))\,\de(u) 
\eeas 
} \hfill \parbox{8mm}{\bea \label{geo}\eea}\\ 
Since  all operations appearing in (\ref{geo}) are well-defined 
in  $\G$ (cf. the remarks following Theorem \ref{geoth} below),   
we  may  seek solutions of the corresponding initial 
value  problem  in  the Colombeau algebra by embedding $\de(u)$ 
into  $\G$.  In  fact,  it turns out that for {\em any} generalized 
delta function there exists a unique solution. Denoting the generalized
functions corresponding to $x^i$ and $v$ by capital letters we state the
following
\begin{theorem} \label{geoth} 
Let   $D\in  \G(\R)$  be  a  generalized  delta function, $f\in 
{\cal C}^\infty(\R^2)$ and let $v_0$, $\dot v_0$, $x^i_0$, 
$\dot x^i_0\in \R$ $(i=1,2)$. The initial value problem \\ 
\parbox{12cm}{ 
\beas
     \ddot V(u)\,&=&f(X^i(u))\,\dot D(u) 
              \,+\,2\,\pa_i f(X^i(u))\,\dot X^i\,(u)\,D(u)\\ 
     \! \ddot X^i(u)&=&\frac{1}{2}\,\pa_i f(X^i(u))\,D(u)\\ 
     V(-1) &=& v_0 \;\;\;\;\;\,  X^i(-1) \;\;=\;\; x^i_0 \\ 
     \dot V(-1) &=& \dot v_0 \;\;\;\;\, \dot X^i(-1)\;\;= \;\;\dot x_0^i 
\eeas 
} \hfill \parbox{8mm}{\bea \label{geocol}\eea} \\ 
has a unique locally bounded solution $(V,X^1,X^2)\in \G(\R)^3$. 
\end{theorem}
Note that we impose initial conditions in $u=-1$, i.e. ``long before''
the shock. Choosing initial conditions at $u=0$ would mean to start ``at the
shock,'' which inevitably leads to regularization dependent weak limits.

\pr   {\em  Existence:}  Choose  $p\in  \N$  as  in  definition 
\ref{gendelta},   fix   $\vphi\in  {\cal  A}_p(\R^n)$  and  let 
$\eps<\eta(\vphi)$.      Then   componentwise   we  obtain  the 
equations\\ 
\parbox{14cm}{ 
\beas 				 
\ddot V(\vphi_\eps,u)&=&f(X^i(\vphi_\eps,u))\dot D(\vphi_\eps,u) 
\,+\,2\,\pa_i f(X^i(\vphi_\eps,u))\dot X^i\,(\vphi_\eps,u)\, 
D(\vphi_\eps,u) \\ 
\ddot X^i(\vphi_\eps,u)& =& \frac{1}{2} 
\pa_i f(X^i(\vphi_\eps,u))\,D(\vphi_\eps,u) \\ 
V(\vphi_\eps,-1) &=& v_0  
\qquad X^i(\vphi_\eps,-1) = x_0^i  \\ 
\dot V(\vphi_\eps,-1)& = &\dot v_0
\qquad \dot X^i(\vphi_\eps,-1) = x_0^i  
\eeas} \hfill  
\parbox{8mm}{\bea \label{geocolrep}\eea} \\ 
According     to    Lemma    \ref{ro},    the  second  line  of 
(\ref{geocolrep})  has  a unique 
globally   defined   solution    $X^i(\vphi_\eps,\,.\,)$  with  the 
specified  initial  values.  Inserting this into the first line and 
integrating  we  also  obtain a solution $V(\vphi_\eps,\,.\,)$. From  
the  boundedness  properties  of  $X^i(\vphi_\eps,\,.\,)$ 
established    in    Lemma   \ref{ro}   and   the   fact   that 
$(D(\vphi,\,.\,))_{\vphi\in{\cal A}_0}\in {\cal E}_M(\R)$ it follows  
easily by induction that $(X^i(\vphi_\eps,\,.\,))_{\vphi\in{\cal A}_0}$  
and  $(V(\vphi_\eps,\,.\,))_{\vphi\in{\cal A}_0}$ are moderate as 
well.   Hence  their  respective  classes  in  $\G(\R)$  define 
solutions to (\ref{geocol}).\\ 
{\em Uniqueness:} Suppose that $V_1=\cl[(V_1(\vphi_\eps, 
\,.\,))_{\vphi\in{\cal A}_0}]$ and $X_1^i =  
\cl[(X^i(\vphi_\eps,\,.\,))_{\vphi\in{\cal      A}_0}]$     are 
locally bounded solutions    of    (\ref{geocol})    as  well.   
On the level of representatives this means that there 
exist $M=\cl[(M(\vphi_\eps,\,.\,))_{\vphi\in{\cal A}_0}], 
\, N^i=\cl[(N^i(\vphi_\eps,\,.\,))_{\vphi\in{\cal A}_0}] \in 
{\cal N}(\R)$ and $n_{x^i}$, $n_{\dot x^i}$, $n_v$, $n_{\dot v}$ $\in 
{\cal  N}(1)$ with\\ 
\parbox{14cm}{ 
\beas 				 
&&\quad\quad\ddot V_1(\vphi_\eps,u)\,\,\,= 
f(X_1^i(\vphi_\eps,u))\,\dot D(\vphi_\eps,u) 
\,+\,2\,\pa_i f(X_1^i(\vphi_\eps,u))\,\dot X_1^i\,(\vphi_\eps,u) \\  
&&\quad\quad\hphantom{\ddot V_1(\vphi_\eps,u)\,\,\,=} \,D(\vphi_\eps,u)+  
M(\vphi_\eps,u) \\ 
&&\quad\quad\ddot X_1^i (\vphi_\eps,u)\;\, = \frac{1}{2} \pa_i  
f(X_1^i(\vphi_\eps,u))D(\vphi_\eps,u) + N^i(\vphi_\eps,u) \\ 
&&\quad\quad V_1(\vphi_\eps,-1) = v_0 + n_v(\vphi_\eps) \;\;\;\; 
X_1^i(\vphi_\eps,-1) = x_0^i + n_{x^i}(\vphi_\eps) \\ 
&& \quad\quad \dot V_1(\vphi_\eps,-1) = \dot v_0 + n_{\dot v}(\vphi_\eps) 
\;\;\;\;							  
\dot X_1^i(\vphi_\eps,-1) = \dot x_0^i + n_{\dot x^i}(\vphi_\eps) 
\eeas} \hfill \parbox{8mm}{\bea \label{geocolrepu}\eea} \\ 
We have to show that $((V-V_1)(\vphi_\eps,\,.\,))_{\vphi\in{\cal A}_0}$ 
and    $((X^i-X_1^i)(\vphi_\eps,\,.\,))_{\vphi\in{\cal   A}_0}$ 
belong  to  the ideal ${\cal N}(\R)$. Since $N^i\in {\cal N}(\R)$ 
it follows that for $p$ sufficiently large, $\eps$ small and 
$\vphi\in {\cal A}_p(\R)$,  
$N^i(\vphi_\eps,\,.\,)$  is bounded on compact sets, uniformly in 
$\eps$. Thus by Lemma \ref{ro} the same holds true for  
$X^i_1(\vphi_\eps,\,.\,)$   and   its   first   derivative.  From 
(\ref{geocolrepu}) we conclude 
\beas 
&& (X^i-X^i_1)(\vphi_\eps,u) = - n_{x^i}(\vphi_\eps) - (u+1) n_{\dot x^i} 
(\vphi_\eps) + \\ 
&&\qquad + \frac{1}{2}\int\limits_{-1}^u\int\limits_{-1}^s D(\vphi_\eps,r) 
[\pa_i  f(X^i(\vphi_\eps,r))  - \pa_i f(X_1^i(\vphi_\eps,r)) ] drds - 
 \int\limits_{-1}^u\int\limits_{-1}^s N(\vphi_\eps,r) dr ds 
\eeas 
Hence  $\forall  T>0$  $\exists  p\in  \N_0$ $\exists \gamma\in 
\Gamma$  $\forall  q\ge  p$  $\forall \vphi\in  {\cal  A}_q(\R)$ 
$\exists C>0$ $\exists  \eta>0$  $\forall \eps\in (0,\eta)$  
$\forall u\in [-T,T]$: \\ 
\parbox{14cm}{ 
\beas 
\quad\quad\quad |(X^i-X^i_1)(\vphi_\eps,u)| \le C\eps^{\gamma(q)-p} +  
\frac{1}{2}\int\limits_{-1}^u\int\limits_{-r}^u  
\int\limits_0^1 
|\nabla\pa_i  f(\sigma X^i(\vphi_\eps,r))+&& \\ 
\quad\quad\quad +(1-\sigma)X^i_1(\vphi_\eps,r))|d\sigma  
|(X^i-X^i_1)(\vphi_\eps,r)||D(\vphi_\eps,r)|dsdr &&  
\eeas} \hfill \parbox{8mm}{\bea \label{Gronwall}\eea} \\ 
By  the  boundedness    properties  of  $X^i$  and  $X_1^i$  and by 
(iii), an application of Gronwall's Lemma to 
the  above  inequality yields the ${\cal N}$-estimates of order 
$0$  for  $(X^i-X^i_1)$.  A  similar  argument applies to the first 
derivatives.   The   estimates  of  higher  order  then  follow 
inductively from the differential equation, so  
$((X^i-X_1^i)(\vphi_\eps,\,.\,))_{\vphi\in{\cal   A}_0}\in 
{\cal N}(\R)$.  Inserting this into the integral equation for 
$(V-V_1)$,   the  ${\cal  N}$-estimates  for  $(V-V_1)$  also  
follow inductively. \epr\medskip\\ 
In  the  proof  of  Theorem  \ref{geoth}  we have only made use 
of  properties  (i)  and (iii) of  
the generalized delta function $D$. On the other hand, property 
(ii)   will  be  essential  for  the  explicit  calculation  of 
distributional  limits  of  the  unique solution constructed in 
Theorem \ref{geoth}, cf. Sec.~\ref{dl}. Also, note that 
we  did  not  have  to impose any growth restrictions on $f$ to 
obtain  a  well-defined  element  $f(X^i)$  of $\G$. This is of 
course  due  to the fact that any componentwise solution of the 
initial  value  problem  necessarily  is  bounded, uniformly in 
$\eps$,    on compact sets (for $\eps$ small). 

Our  next  goal  is  an  analysis  of  the  Jacobi equation for 
impulsive  pp-waves in the framework of algebras of generalized 
functions.   As   in   \cite{geo} to keep formulas more transparent    
we  make  some  simplifying 
assumptions  concerning  geometry (namely axisymmetry) and 
initial  conditions. Writing $x=x^1$ and $y=x^2$ we  suppose  
that  $f$ depends
exclusively  on the two-radius $\sqrt{x^2+y^2}$ and work within 
the hypersurface $y=0$ (corresponding to initial conditions  
$y_0=0=\dot  y_0$).  Furthermore we demand 
$v_0=0=\dot x_0$. As was shown in \cite{geo}, in this 
situation the Jacobi equation 
\[ 
\frac{D^2N^a}{dt^2}=-R\,^a_{bcd}T^b T^d N^c\,\,, 
\] 
where $N^a(u)=(N^u(u),N^v(u),N^x(u),N^y(u))$ denotes the 
deviation vector field takes the form 
\\ 
\parbox{14cm}{ 
\beas 
     \ddot N^v &=&2[N^x f'(x )\de]\,\dot{\,}- 
                        N^x f'(x )\dot\de + 
                        [N^u f(x )\de]\,\ddot{\,}
			- N^u f''(x )\dot x^2 \de - 
                        N^u f'(x )\ddot x \de \\ 
     \ddot N^x &=&[\dot {N^u} f'(x )+ 
                         \fr{2}\,N^x f''(x )]\de  
                         +\fr{2}\,f'(x )N^u \dot\de \\ 
     \ddot N^y &=& \ddot N^u \;=\; 0\,\,, 
\eeas} \hfill \parbox{10mm}{\bea \label{jacobi}\eea} \\ 
where $x$ is determined by (\ref{geo}).  Existence  and  
uniqueness of solutions to the    corresponding  initial  value   
problem  in the Colombeau algebra is established in the following  
result where, for the sake of brevity we denote the $\G$-functions 
corresponding to $N^a$ again by $N^a$.

\begin{theorem} \label{jacth} 
Let   $D\in  \G(\R)$  be  a  generalized  delta function, $f\in 
{\cal C}^\infty(\R)$, $n^a$, $\dot n^a\in  \R^4$ and let $X$ denote  
the (unique) solution to system (\ref{geocol}) with initial conditions 
and simplifications as discussed above.  
The initial value problem \\ 
\parbox{12cm}{ 
\beas
       \quad   \ddot N^v &=&2[N^x f'(X)D]\,\dot{\,}- 
                        N^x f'(X)\dot D + 
                        [N^u f(X) D]\,\ddot{\,} - \\ 
     \quad\hphantom{N^v} &\hphantom{=}&                    
			- N^u f''(X)\dot X^2 D - 
                        N^uf'(X)\ddot X D \\ 
     \quad\ddot N^x &=&[\dot {N^u} f'(X)+ 
                         \fr{2}\,N^x f''(X)] D  
                         +\fr{2}\,f'(X)N^u \dot D \\ 
     \quad\ddot N^y &=& \ddot N^u \;=\; 0\\ 
     \quad N^a(-1) &=& n^a \;\;\;\;\;\,  \dot N^a(-1) \;=\; \dot n^a \\ 
\eeas}  
\hfill \parbox{10mm}{\bea \label{jacol}\eea} \\ 
has a unique solution $N^a\in \G(\R)^4$. 
\end{theorem}  
\pr Since the equations are linear in the components of the deviation field 
we are provided with globally defined solutions on the level of representatives. 
The last two equations are actually trivial and so is the first one once we know 
that its right hand side belongs to $\G(\R)$. Hence we are left 
with the equation for $N^x$ which is of  the form  
$\ddot N(t) = f''(X(t))D(t)N(t) + H(t)$ with $H$ in $\G(\R)$.  
Using the boundedness properties of $X$ established in 
Lemma~\ref{ro} the $\E_M$-bounds for $N^x$ easily follow from Gronwall's lemma. 
 
Uniqueness is established along the same lines again using Gronwall-type 
arguments.  
\epr
 
In  the  above  proof  we  have  again  only  used  properties  
(i)  and (iii) of the generalized delta function $D$. 

To  conclude  this  section  we remark that unique solvability of 
the geodesic and geodesic deviation equation for (\ref{metric}) 
is  not  confined  to  the case where the profile function $f$  
is  smooth.  Indeed,  it  turns  out  that for a large class of 
generalized   profile   functions  (those  that  are  not  ``too
singular'') Theorems \ref{geoth} and  
\ref{jacth} retain their validity. More precisely, we have to demand that $f$
belongs to the algebra of {\em tempered} generalized functions~\cite{co2} to
make sure that the composition $f(X)$ is well defined and that $\nabla\nabla f$
is of $L^\infty$-log-type~\cite{moODE,kunzDISS} to ensure existence and
uniqueness of solutions to (\ref{geocol}) and (\ref{jacol}). 
However, to include many
physically interesting examples (cf.~\cite{EX}) one has to cut out the 
worldline of the ultrarelativistic particle, i.e. the $v$-axis from the 
domain of definition (cf.~\cite{stein}).
\section{Distributional Limits}\label{dl}  
In  this  section  we are going to calculate the distributional 
limits  (or,  in  the  terminology  of  Colombeau  theory:  the 
associated  distributions)  of  the  unique  solutions  to  the 
geodesic   and   geodesic  deviation  equation  constructed  in 
Theorems    \ref{geoth}    and   \ref{jacth}.   
In~\cite{geo}
distributional   limits   for  regularized  versions  of  these 
equations   have   been   calculated using a model  delta  net 
regularization.
Translated  into our current setting this 
amounts  to using the particular generalized delta function $D= 
\iota(\delta)$.  Our  aim  is to extend the validity of the 
limit    relations  derived  there to the case of solutions in 
the  Colombeau  algebra  and to generalized delta functions. At 
the     same    time    we    will  be  able  to  prove  
stronger convergence results in some cases. 

\begin{theorem} \label{geoass} 
The   unique   solution  $(V,X^i)$  of  the  geodesic  equation 
(\ref{geocol})    satisfies    the    following    association 
relations: 
\bea 
X^i &\approx & x_0^i + \dot x_0^i (1+u) + \frac{1}{2}\pa_i  
f(x_0^i + \dot x_0^i) u_+ \label{Xass}\\ 
V\, &\approx & v_0 + \dot v_0 (1+u) + f(x_0^i + \dot x_0^i) \theta(u) 
+\pa_if(x_0^i + \dot x_0^i) \left(\dot x_0^i +  
\frac{1}{4}\pa_if(x_0^i +  \dot x_0^i)\right) u_+ \label{Vass} 
\eea 
In  addition,  if  $X^i = \cl[(X^i(\vphi,\,.\,))_{\vphi\in {\cal 
A}_0}]$ then $\exists p\in \N_0$ such that $\forall \vphi\in{\cal A}_p$ 
\beq \label{Xunif} 
X^i(\vphi_\eps,u) \to x_0^i + \dot x_0^i (1+u) + \frac{1}{2}\pa_i  
f(x_0^i + \dot x_0^i) u_+  
\eeq 
for $\eps\to 0$, uniformly on compact subsets of $\R$. 
\end{theorem} 
 
\pr Choose $p\in \N_0$ as in definition \ref{gendelta} for $D$, 
let  $\vphi  \in  {\cal  A}_p$  and  $\eps<\eta(\vphi)$.  Since 
integrating  amounts  to convolution with the Heaviside function,
which  is  a continuous operation on the convolution algebra of 
distributions   supported   in   a  cone,  in  order  to  prove 
(\ref{Xass}) it suffices to show that 
\[ 
\ddot X^i(\vphi_\eps,\,.\,) = \frac{1}{2}\pa_i f(X^i(\vphi_\eps,\,.\,))  
D(\vphi_\eps,\,.\,)\to \frac{1}{2}\pa_i f(x_0^i + \dot x_0^i) \de 
\] 
in distributions. We first note that $X(\vphi_\eps,\eps t)\to 
x_0^i  +  \dot x_0^i$ uniformly as can be seen from the integral equation 
for $X^i$ (cf. (11) in \cite{geo}). Now if $\psi \in \DD(\R)$  
then 
\beas 
&&\mid\int\limits_{-\ep}^\ep\psi(t)\pa_if(X^i(\vphi_\ep,t))D(\vphi_\ep,t)dt-
	\pa_if(x^i_0+\dot x^i_0)\psi(0)\mid\nn\\ 
&&\quad\leq \sup_{-\ep\leq t\leq \ep}\mid\psi(t)\pa_if(X^i(\vphi_\ep, 
	t))-\pa_if(x_0^i+\dot x^i_0)\psi(0)\mid\int\limits_ 
	{-\ep}^\ep\mid D(\vphi_\ep,t)\mid dt +\nn\\ 
&&\quad +\int\limits_{-\ep}^\ep\mid D(\vphi_\ep,t)dt-1\mid 
	\pa_if(x_0^i+\dot x_0^i)\psi(0) 
\eeas 
So the claim follows from properties (iii) and (ii) of the generalized  
delta function $D$. 
Since $\dot X^i(\vphi_\eps,t)$ 
is bounded on compact sets, uniformly in $\eps$, it follows that 
the   family   $\{X^i(\vphi_\eps,t):\,  \eps\in  (0,1)\}$  is 
locally  equicontinuous. Hence Ascoli's Theorem implies  
(\ref{Xunif}).   Concerning  (\ref{Vass}), as above it suffices 
to calculate the limit of  
\[ 
\ddot V(\vphi_\eps,u)=[f(X^i(\vphi_\eps,u)) D(\vphi_\eps,u)] 
\dot{\,} +\pa_i f(X^i(\vphi_\eps,u))\dot X^i\,(\vphi_\eps,u) 
D(\vphi_\eps,u) 
\] 
whose first summand converges to $f(x_0^i + \dot x_0^i)\dot\de$ 
by  an argument similar to the one above. For the second summand we 
have 
\beas 
&&\pa_i f(X^i(\vphi_\eps,u)) \dot X^i(\vphi_\eps,u) D(\vphi_\eps,u) 
= \underbrace{\pa_i f(X^i(\vphi_\eps,u)) D(\vphi_\eps,u)  
\dot x_0^i}_{(\ast)} \\ 
&& +\frac{1}{2}\pa_i f(X^i(\vphi_\eps,u)) D(\vphi_\eps,u) 
\int\limits_{-\eps}^t \pa_i f(X^i(\vphi_\eps,s))D(\vphi_\eps,s) ds 
\eeas 
and $(\ast) \to \pa_i f(x_0^i + \dot x_0^i)\dot x_0^i \de$. 
Finally, since 
\beas 
&& \int\limits_{-\eps}^\eps \psi(t) \pa_i  
f(X^i(\vphi_\eps,t)) D(\vphi_\eps,t) 
\int\limits_{-\eps}^t \pa_i f(X^i(\vphi_\eps,s))D(\vphi_\eps,s) 
dsdt -\\ 
&& -\frac{1}{2}\pa_i f(x_0^i + \dot x_0^i)^2\psi(0)\int\limits_ 
{-\eps}^\eps D(\vphi_\eps,t)dt \to 0 \, , 
\eeas
the claim follows. \epr \medskip 

In calculating  distributional limits for the solution of 
the Jacobi equation to maintain clarity of formulae we shall
make simplifying assumptions on the initial conditions, i.e. 
\parbox{12cm}{ 
\beas 
N^a (-1) &=& (0,0,0,0) \\ 
\dot N^a (-1) &=& (a,b,0,0)\,\,. 
\eeas} \hfill \parbox{10mm}{\bea \label{NIC}\eea} \\ 
Then we have 
\begin{theorem} \label{jacass} 
The   unique   solution   of  the  geodesic deviation equation 
(\ref{jacol})    satisfies    the    following    association 
relations: 
\bea 
N^x &\approx & \frac{1}{2} a f'(x_0)(u_++\theta(u)) \label{nxass}\\ 
N^v &\approx & b(1+u) + a[f(x_0)\de(u) +\frac{1}{4}f'(x_0)^2 
(\theta(u)+u_+)] \label{nvass} 
\eea 
\end{theorem}

\pr  The general structure of this proof is `isomorphic' to the 
calculation of distributional limits for the regularized Jacobi 
equation in \cite{geo}. The main difference is that for 
representatives   of   generalized  delta  functions  dominated 
convergence  arguments  are  not  applicable  which  makes  the 
calculations  more  tedious. Nevertheless, using the uniform convergence 
of $X^i(\vphi_\ep,.)$ established above, all steps carried out 
in  \cite{geo} can be adapted to the present situation  
as  demonstrated in the proof of Theorem \ref{geoass}. \epr\medskip

\section{Discussion and outlook}\label{mf} 

In the previous section we have 
shown that the unique solutions to the geodesic 
and geodesic deviation equation in the Colombeau algebra
possess a physically reasonable macroscopic (i.e. distributional)  
aspect:   even   within   the   natural   maximal   class   of 
delta-regularizations  (namely  the  class  of  all generalized 
delta  functions)  the  regularity  of the equations is 
sufficiently high to ensure distributional limits corresponding 
to physical expectations. 
More precisely, from the distributional point of view,
the geodesics correspond to refracted, broken straight lines as suggested by 
the form of the metric. The scale of the jump and kink is given by 
the values of $f$ and its first derivatives at the shock hypersurface, 
which can be traced back to the values at the initial point ($u=-1$), thereby
precisely reproducing Penrose's junction conditions~\cite{penrose}.
The distributional limit of the Jacobi field suffers a
kink and jump in $x$-direction as well as an additional $\de$-pulse in
$v$-direction, which may be understood from the form of the geodesics. For a
more detailed discussion see~\cite{geo}.

Finally we make some comments on diffeomorphism invariance of our results.
Whereas the fine sheaf of Colombeau algebras can be lifted to manifolds 
in a straightforward manner, the action of a diffeomorphism does not commute
with the canonical embedding ${\cal D}'\hookrightarrow\G$. The 
reason  for  this  is that convolution relies on the additive 
group  structure of $\R^n$ and is therefore not invariant under 
the action  of  diffeomorphisms. Note, however that our calculations did not use
the embedding via convolution and therefore are not affected by this defect.

A solution to the above mentioned problem  was proposed in~\cite{CM} 
using a modified definition of the mollifier spaces ${\cal A}_q$. 
A key ingredient of this construction is that diffeomorphisms act on 
the $\vphi$'s, introducing an implicit $x$-dependence into the first slot of the 
Colombeau functions $R(\vphi,x)$. Hence, to retain smooth $x$-dependence of 
$R$ the concept of Silva-differentiability was used. Future work will be
concerned with a simplified concept of Colombeau algebras on manifolds using
calculus in convenient vector spaces~\cite{KM}.
A main goal of this line of research is to provide a workable
solution concept for singular differential equations on manifolds.

\section*{Acknowledgement} 
The authors wish to thank M. Grosser, M. Oberguggenberger and  
H. Urbantke for helpful discussions. 
M. Kunzinger was supported by Research Grant P10472-MAT of the Austrian  
Science Foundation. R. Steinbauer was supported by Austrian Academy of  
Science, Ph.D. programme, grant \#338 and by Research Grant P12023-MAT  
of the Austrian Science Foundation. 
                 

\begin{thebibliography}{99} 
\bibitem{penrose}
	R. Penrose,
  	``The geometry of impulsive gravitational waves,''
	in {\it General Relativity, Papers in Honour of J.\,L. Synge,}
  	edited by L. O'Raifeartaigh (Clarendon Press, Oxford, 1972), 
	pp. 101-115.
\bibitem{as}  
	P.\,C. Aichelburg, R.\,U. Sexl,  
	``On the gravitational field of a massless particle,'' 
	J. Gen. Rel. Grav. {\bf 2}, 303-312 (1971). 
\bibitem{bn} 
	H. Balasin, H. Nachbagauer,  
	``The ultrarelativistic Kerr-geometry and its energy-momentum tensor,'' 
	Class. Quant. Grav. {\bf 12}, 707-713 (1995). 
\bibitem{ls}
	C.\,O. Loust\'o, N. S\'anchez, 
	``The ultrarelativistic limit of the Kerr-Newman geometry and 
	particle scattering at the Planck scale,''
	Phys. Lett. B {\bf 232}, 462-466 (1989).
\bibitem{pg}
	J. Podolosky, J.\,B. Griffiths,  ``Boosted static multipole particles as 
	sources of impulsive gravitational waves,'' to appear in Phys. Rev.
	D., gr-qc/9809003, (1998).   
\bibitem{vv}
	E. Verlinde, H. Verlinde, 
	``High-energy scattering in Quantum Gravity,'' 
	Class. Quant. Grav. {\bf 10}, Suppl. 175-184 (1993). 
\bibitem{dt} 
	T. Dray, G. 't Hooft,  
	``The gravitational shockwave of a massless particle,''
	Nucl. Phys. B {\bf 232}, 173-188 (1985). 
\bibitem{aib}P.\,C. Aichelburg, H. Balasin, ``Generalized symmetries of
        impulsive gravitational waves,'' Class. Quant. Grav. {\bf 14},
        A31-A41, (1997).    
\bibitem{ba} 
	H. Balasin, 
	``Geodesics for impulsive gravitational waves and the multiplication
	 of distributions,'' 
	Class. Quant. Grav. {\bf 14}, 455-462 (1997). 
\bibitem{fvp} 
	V. Ferrari, P. Pendenza, G. Veneziano, 
	``Beam like gravitational waves and their geodesics,''
	J. Gen. Rel. Grav. {\bf 20}, 1185-1191 (1988). 
\bibitem{vp} 
	J. Podolsky, K. Vesely, ``Continuous coordinates for all 
  	impulsive pp-waves,'' Phys. Lett. A241, 145-147 (1998).
\bibitem{geo} 
 	R. Steinbauer, ``Geodesics and geodesic deviation for impulsive
  	gravitational waves,'' J. Math. Phys. {\bf 39}, 2201-2212 (1998).
\bibitem{hajek} 
	O. H\'ajek, Bull. AMS {\bf 12}, 272-279 (1985).
\bibitem{co1} 
	J.\,F. Colombeau,  
	``New Generalized Functions and Multiplication of Distributions,''   
	(North Holland, Amsterdam, 1984). 
\bibitem{co2} 
	J.\,F.Colombeau, 
	``Elementary Introduction to New Generalized
	Functions,'' (North Holland, Amsterdam 1985).
\bibitem{mo} 
	M. Oberguggenberger, 
	``Multiplication of Distributions and Applica\-tions to Partial 
	Differential Equations,'' 
	(Longman Scientific and Technical, New York, 1992).
\bibitem{moODE} 
	R. Hermann, M. Oberguggenberger,
	``Generalized functions, calculus of variations and nonlinear ODEs,''
	to appear in {\it Nonlinear Theory of Generalized Functions,}
	edited by 	
	M. Grosser, G. H\"ormann, M. Kunzinger, M.Oberguggenberger,
 	(Pitman Research Notes in Mathematics, 1999). 
\bibitem{kunzDISS}
	M. Kunzinger,
	``Lie Transformation Groups in Colombeau Algebras,''
	Ph.D.-thesis, University of Vienna (1996).
\bibitem{pv2}J. Podolsky, K. Vesely,
        ``New examples of sandwich gravitational waves and their impulsive
        limit,'' Czech. J. Phys. {\bf 48}, 871, (1998).
\bibitem{AB} 
        J. Aragona, H. A. Biagioni, ``Intrinsic definition   
	of the Colombeau algebra of generalized functions,''   
	{\it Analysis Mathematica}, {\bf 17}, 75 - 132  (1991). 
\bibitem{EX}
	C.\,O. Loust\'o, N. S\'anchez, 
	``Gravitational shock waves generated by extended sources:
	Ultrarelativistic cosmic strings, monopoles and domain walls,''
	Nucl. Phys. B {\bf 355}, 231-249 (1991).
\bibitem{stein} 
	R. Steinbauer, 
   	``The  ultrareltivistic  Reissner-Nordstr\o  m  field in the 
   	Co\-lom\-beau algebra,'' 
   	J. Math. Phys. {\bf 38}, 1614-1622 (1997). 
\bibitem{CM} 
	J.\,F. Colombeau, A. Meril, ``Generalized functions
	and   multiplication   of   distributions   on   manifolds,'' 
	J. Math. Anal. Appl. {\bf 186}, 357-364 (1994). 
\bibitem{KM} 
	A. Kriegl,  P.\,W. Michor,    
	``The  Convenient  Setting of Global Analysis,'' 
	(Mathematical Surveys and Monographs {\bf 53}, AMS 1997).

\end{thebibliography}
\end{document}